\documentclass[journal,twoside,web]{ieeecolor}
\usepackage{generic}
\usepackage{cite}
\usepackage{amsmath,amssymb,amsfonts}
\usepackage{bbold}
\usepackage{graphicx}
\usepackage{textcomp}
\usepackage{soul, color}
\usepackage{steinmetz}
\usepackage{comment}
\usepackage{algorithm2e}
\SetKwComment{Comment}{/* }{ */}
\SetKwInput{kwPreparation}{Preprocessing}
\SetKwInput{kwTraining}{Training}
\SetKwInput{kwInference}{Evaluation}
\RestyleAlgo{ruled}
{\begin{list}{$\bullet$ \hfill}{
			\setlength{\leftmargin}{\parindent}
			\setlength{\parsep}{0.07\baselineskip}
			\setlength{\itemsep}{0.7\parsep}
			\setlength{\labelwidth}{\leftmargin}
			\setlength{\labelsep}{0em}}}{\end{list}}
\def\BibTeX{{\rm B\kern-.05em{\sc i\kern-.025em b}\kern-.08em
    T\kern-.1667em\lower.7ex\hbox{E}\kern-.125emX}}
\begin{document}
\title{Machine-to-Machine Transfer Function in Deep Learning-Based Quantitative Ultrasound}
\author{Ufuk Soylu, \textit{Student Member, IEEE} and Michael L. Oelze, \textit{Senior Member, IEEE}
\thanks{This research received financial support from grants provided by the National Institutes of Health (NIH) (R01CA251939 and R01CA273700)}
\thanks{Ufuk Soylu and Michael Oelze are with the Beckman Institute, and the Department of Electrical and Computer Engineering, University of Illinois at Urbana-Champaign, Urbana, IL 61801 USA. Moreover, Michael Oelze is with the Carle Illinois College of Medicine, University of Illinois at Urbana-Champaign, Urbana, IL 61801 USA. (e-mail: usoylu2@illinois.edu; oelze@illinois.edu).}
}
\maketitle
\begin{abstract}
A Transfer Function approach was recently demonstrated to mitigate data mismatches at the acquisition level for a single ultrasound scanner in deep learning (DL) based quantitative ultrasound (QUS). As a natural progression, we further investigate the transfer function approach and introduce a Machine-to-Machine (M2M) Transfer Function, which possesses the ability to mitigate data mismatches at a machine level, i.e., mismatches between two scanners over the same frequency band. This ability opens the door to unprecedented opportunities for reducing DL model development costs, enabling the combination of data from multiple sources or scanners, or facilitating the transfer of DL models between machines with ease. We tested the proposed method utilizing a SonixOne machine and a Verasonics machine. In the experiments, we used a L9-4 array and conducted two types of acquisitions to obtain calibration data: stable and free-hand, using two different calibration phantoms. Without the proposed calibration method, the mean classification accuracy when applying a model on data acquired from one system to data acquired from another system was approximately 50\%, and the mean AUC was about 0.40. With the proposed method, mean accuracy increased to approximately 90\%, and the AUC rose to the 0.99. Additional observations include that shifts in statistics for the z-score normalization had a significant impact on performance. Furthermore, the choice of the calibration phantom played an important role in the proposed method. Additionally, robust implementation inspired by Wiener filtering provided an effective method for transferring the domain from one machine to another machine, and it can succeed using just a single calibration view without the need for multiple independent calibration frames.

\end{abstract}
\section{Introduction}
\label{sec:intro}

In the realm of biomedical ultrasound imaging, deep learning (DL) holds great potential for advancing the field, driven by significant interest from both academia and industry. As DL models become more sophisticated, large datasets become increasingly available, and computational power scales up, the capability of DL to address clinical tasks gets closer to integration into clinical workflows, potentially leading to a transformation in the field of ultrasound imaging. At its essence, DL algorithms learns a sequence of nonlinear transformations, each customizable with parameters, used to derive multiple layers of features from input image data and then make predictions in an automated way, eliminating the need for manual feature extraction. DL is capable of learning high-dimensional functional approximations to perform complex desired behaviors, seemingly defying the curse of dimensionality. Convolutional neural networks (CNNs) emerged as the most preferred and studied approach among DL algorithms in ultrasound biomedical imaging due to their efficiency in analyzing images\cite{liu2019deep}.

The adoption of DL-powered biomedical ultrasound imaging in clinical settings, in an ethical, interpretable, and trustworthy way, is the coveted goal of both industry and the research academy. DL-powered biomedical ultrasound can lead to a significant increase in the quality of medical services in an automated and efficient manner. However, achieving this goal requires major breakthroughs in DL algorithms. Two main technical challenges hinder the implementation of DL-driven algorithms in actual clinical environments \cite{castro2020causality}. First, for a particular domain there is often a shortage of labeled data, largely because of the high costs associated with conducting laboratory experiments or obtaining expert annotations from clinical data. Second, the issue of data mismatch arises when the conditions in which a DL model is developed differ from those it will face in clinical setting, which can limit the model's ability to generalize effectively. In situations where labeled data is scarce or major differences exist between development and deployment environments, any DL-based algorithm might yield poor clinical performance. Consequently, enhancing the data efficacy and robustness of DL algorithms stands as a crucial research direction for establishing DL as a viable tool in ultrasound imaging.

Similar to the general trend in medical imaging, Quantitative Ultrasound (QUS) has transitioned from classical approaches that rely on manual feature engineering, statistical assumptions, and ad hoc models to DL-based approaches that rely on an abundance of big data and the assumption that training and testing data distributions are identical. Specifically, in several recent examples, CNNs were used to classify tissue states, and it was shown that they outperformed traditional QUS approaches \cite{byra2022joint, byra2022liver,han2020noninvasive, soylu2022data, nguyen2019reference, nguyen2021use}. Following this, in our previous work, a transfer function approach was developed using a calibration phantom to mitigate acquisition-related data mismatches within the same imaging machine for DL-based QUS approaches \cite{soylu2023calibrating}. The transfer function approach significantly improved mean classification accuracies for pulse frequency, output power, and focal region mismatches within the same imaging machine, increasing them from 52\%, 84\%, and 85\% to 96\%, 96\%, and 98\%, respectively. Therefore, the transfer function approach has emerged as an economical way to generalize a DL model for tissue characterization in cases where scanner settings cannot be fixed, thus improving the robustness of DL-based algorithms.

There is a wide and rich literature anthology related to the data mismatch problem in DL \cite{DBLP:journals/corr/abs-2110-11328, DBLP:journals/corr/abs-2007-01434}. For example, data augmentation is a crucial tool for minimizing data mismatch. Some approaches build heuristic data augmentations to approximate the distribution shift between testing and training data, aiming to improve robustness \cite{DBLP:journals/corr/HeZRS15, hendrycks2020augmix, DBLP:journals/corr/abs-1909-13719, cubuk2019autoaugment}. The performance of these approaches depends on how well the approximation mitigates the distribution shift. Other approaches attempt to learn data augmentation by training a generative model between testing and training domains \cite{goel2020model,choi2018stargan,zhu2017unpaired}. On the other hand, domain generalization approaches aim to recover feature representations that are independent of domains \cite{arjovsky2019invariant, sun2016deep, ganin2016domain}. Their performance relies on the invariance of the learned features. Additionally, BN-Adapt \cite{schneider2020improving} modifies batch normalization layers adaptively using test domain data. Moreover, pretraining is another significant concept \cite{sharif2014cnn, azizpour2015generic}. Pretraining on a larger dataset could provide robust representations for downstream tasks.

The issue of data mismatch has gained increased attention in recent literature focusing on DL-based QUS \cite{tehrani2022robust, kim2021learning,tierney2020domain, tehrani2021ultrasound, soylu2023calibrating, sharifzadeh2021ultrasound}. Adaptive batch normalization was utilized in the context of DL-based QUS \cite{tehrani2022robust}. Additionally, cycle-consistent generative adversarial networks were applied to address the issue of data mismatches in ultrasound imaging \cite{tierney2020domain}. Furthermore, the Fourier Domain Adaptation technique was employed, proposing the replacement of lower frequency components within the frequency spectrum \cite{sharifzadeh2021ultrasound}. In contrast to these methodologies, the transfer function approach developed in \cite{soylu2023calibrating} does not require sample data from testing domain. Instead, it relies on a calibration phantom that can be tailored to the specific characteristics of sample data at hand. To the best of our knowledge, the transfer function approach is the only method in the literature that does not rely on real sample data and requires only a calibration phantom, positioning it as a practical method for clinical setting. 

As the transfer function holds the potential for practical implementation within clinical settings, given that it does not necessarily require real samples from the testing domain, it is essential to further validate and identify its strengths and weaknesses under more substantial mismatches. In this study, the application of the transfer function approach was extended to address data mismatches between different imaging machines. By doing so, the transfer function approach would increase its utility in multiple ways. First, being able to transfer between machine domains can lower the cost of DL-based QUS approaches. Specifically, data from different machines can be combined to develop more robust and accurate DL-based models. This has the potential to provide a simple and efficient means of utilizing existing data from different machines and sources, which helps address the high cost associated with labelled data collection. Additionally, DL-based QUS approaches which are developed for specific machines, can be transferred to other machines at ease. Overall, in our prior work, we demonstrated that the transfer function approach has potential to provide an economical way to provide in-system transferability \cite{soylu2023calibrating}. In this work, we demonstrate that transfer functions can be defined that can also provide out-system transferability, i.e., a Machine-to-Machine (M2M) transfer function. Further details of our methodology and experimental results can be found in Sections \ref{sec:met} and \ref{sec:results}, respectively. We then provide a section on discussion related to the research findings in Section \ref{sec:discussion} and conclusions in Section \ref{sec:conclusion}.

\section{Methods}
\label{sec:met}

\subsection{Phantoms}

\begin{figure}[hbt!]
\begingroup
    \centering
    \begin{tabular}{c c }
\hspace{-0.25cm}{ \includegraphics[scale=0.034]{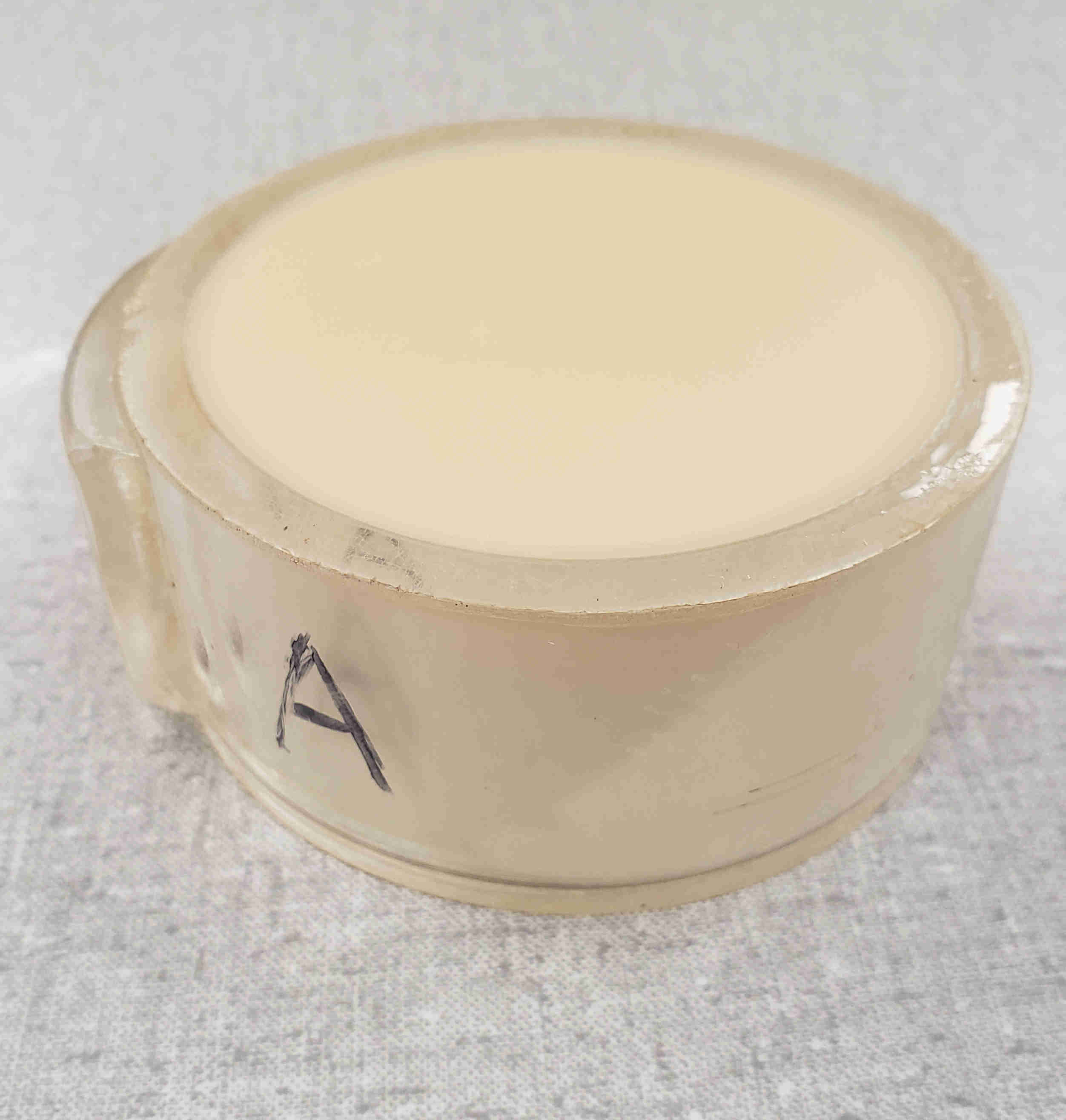}}&\hspace{-0.25cm}{ \includegraphics[scale=0.0326]{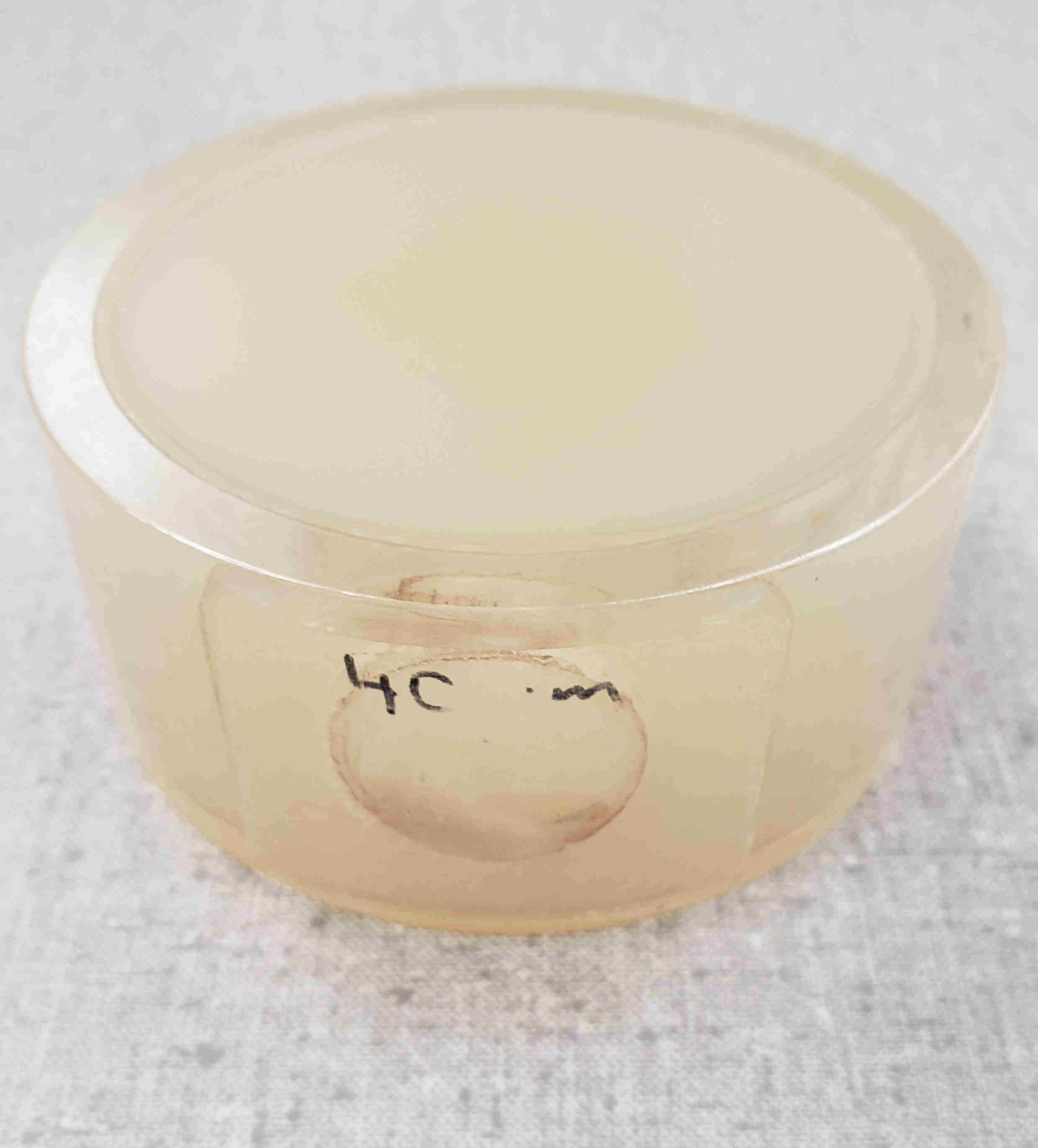}} \\
\hspace{-0.25cm} Classification Phantom 1 &\hspace{-0.25cm} Classification Phantom 2 
\end{tabular}
\caption{Visuals of The Classification Phantoms}
\label{fig:phantom_picutes}
\endgroup
\end{figure}

\begin{figure}[hbt!]
\begingroup
    \centering
    \begin{tabular}{c c}
\hspace{-0.25cm}{ \includegraphics[scale=0.028]{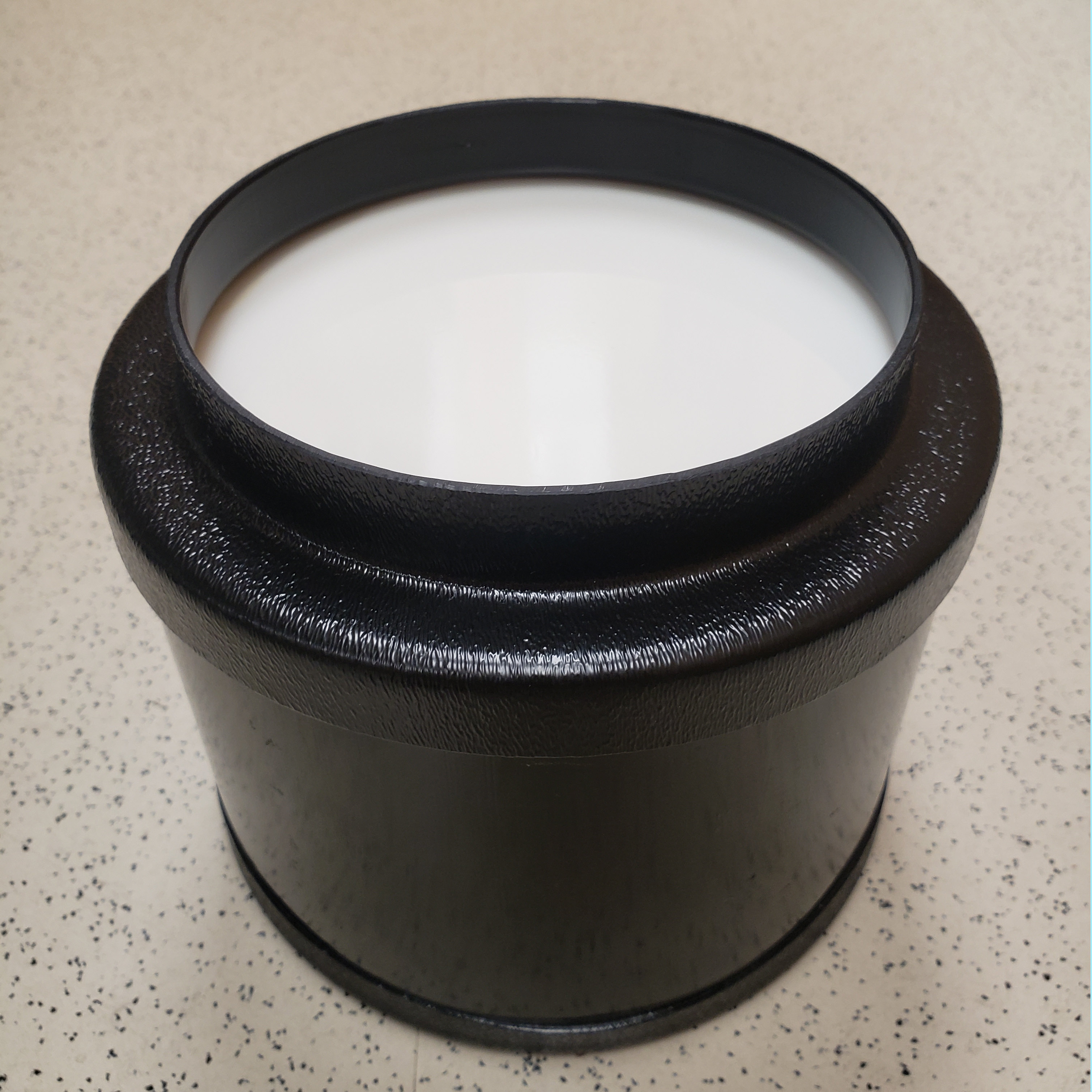}}&\hspace{-0.25cm}{ \includegraphics[scale=0.0485]{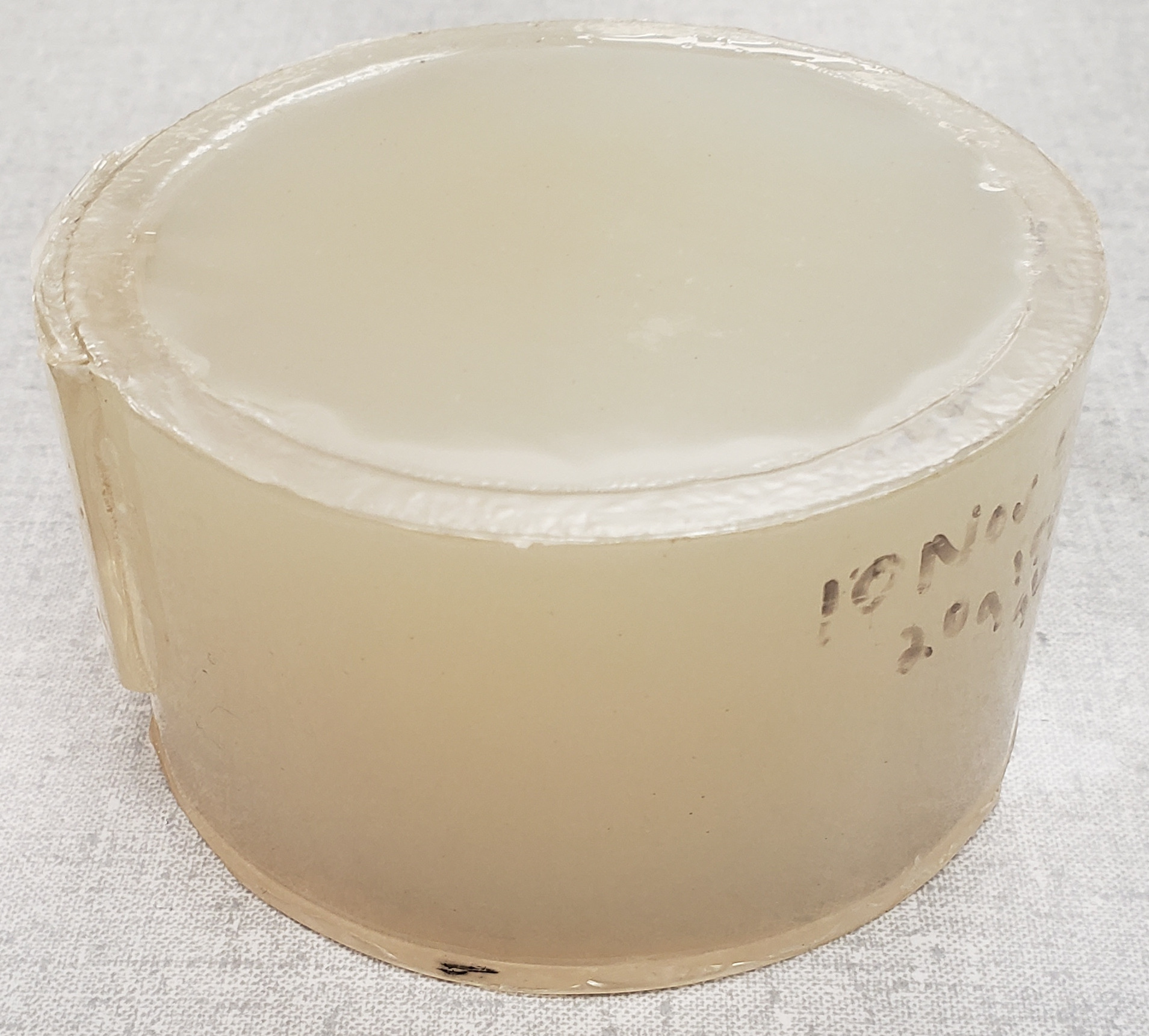}}\\
\hspace{-0.25cm} Calibration Phantom 1 &\hspace{-0.25cm} Calibration Phantom 2 \\
\end{tabular}
\caption{Visuals of The Calibration Phantoms}
\label{fig:calib_picutes}
\endgroup
\end{figure}

The experiments utilized two distinct tissue-mimicking phantoms as classification phantoms, shown in Fig. \ref{fig:phantom_picutes}. Additionally, two distinct calibration phantoms were used to obtain the M2M transfer function, shown in Fig. \ref{fig:calib_picutes}.

Classification Phantom 1 mimics the characteristics of the human liver \cite{wear2005interlaboratory} and the construction details were given in \cite{madsen1998liquid}. The attenuation coefficient slope for Classification Phantom 1 was measured around 0.4 dB$\times$cm$^{-1}\times$MHz $^{-1}$. It exhibited macroscopic uniformity. The sole source of non-uniformity in Classification Phantom 1 stemmed from the random distribution of microscopic glass bead scatterers, ranging in diameter from 75 to 90 $\mu$m. 

Classification Phantom 2 was characterized as a low-attenuation phantom \cite{anderson2010interlaboratory} and the construction details were given in \cite{madsen1978tissue}. The same weakly-scattering agar, serving as the background material, was utilized in Classification Phantom 2 but included glass-bead scatterers of varying sizes, ranging from 39 to 43 $\mu$m in diameter. Its speed of sound was 1539 $m \times s^{-1}$. The attenuation coefficient slope measured around 0.1 dB$\times$cm$^{-1}\times$MHz $^{-1}$. 

The Calibration Phantom 1 was a commercial QUS reference phantom (part no. 14090502, serial no. 221447541) from CIRS, Inc., Norfolk, VA. It had an attenuation coefficient slope of approximately equal to 0.74 dB$\times$cm$^{-1}\times$MHz $^{-1}$. Its speed of sound was 1545 $m \times s^{-1}$. 

The Calibration Phantom 2 was characterized as a low-attenuation phantom \cite{anderson2010interlaboratory}. It was constructed with a 2\% agar background having weakly scattering properties. This phantom included glass beads with diameters measuring 160 $\pm$ 60 $\mu$m. The distribution of glass beads, occurring spatially randomly within the phantom's volume, was at a concentration of 20 g/L. The attenuation coefficient slope for Classification Phantom 2 measured approximately 0.6 dB$\times$cm$^{-1}\times$MHz $^{-1}$. Its speed of sound was 1535 $m \times s^{-1}$.

\subsection{Ultrasound Machines}
The phantoms were scanned using both a SonixOne system and a Verasonics Vantage 128. An L9-4/38 transducer was utilized throughout the experiments. The SonixOne system captured post-beamformed radio-frequency (RF) data. Its sampling rate was 40 MHz. On the other hand, the Verasonics system acquired raw channel data at a sampling rate of 50 MHz. Following that, delay and sum beamforming was implemented on the Verasonics data. Following that, a multirate FIR filter was designed with an interpolation factor of 4 and a decimation factor of 5 to convert the sampling rate to 40 MHz. Beamforming and sampling rate conversion were implemented using MATLAB functions. Specifically, the 'designMultirateFIR' function was used, which computes the filter coefficients based on the interpolation and decimation factors, while the 'dsp.FIRRateConverter' function was used to implement a combined anti-aliasing FIR filter using these filter coefficients, the decimation factor, and the interpolation factor. After these preprocessing steps, post-beamformed RF data at a matching sampling frequency of 40 MHz was obtained from both machines for DL operations. Matching the sampling rate between systems was critical to being able to implement the M2M transfer function.

As training data, the SonixOne data was utilized during the experiments, positioning the SonixOne as the "training machine" where the model development occurred. On the other hand, the Verasonics data was utilized as testing data during the experiments, positioning it as the "testing machine" where the machine data is assumed to be unavailable during model development. Testing machine data was only used to measure calibration success during inference time. For training data and testing data, free-hand data acquisition was utilized with Classification Phantom 1 and Classification Phantom 2, i.e., the transducer was moved across the phantom surface by hand. During this acquisition, by recording a video of 1000 frames, we captured a large amount of ultrasound data for each phantom.

For calibration data, both the SonixOne and Verasonic machines were utilized in two scanning procedures using Calibration Phantom 1 and Calibration Phantom 2. In the first procedure, similar to the training and testing data, free-hand acquisition was utilized, which provided 1000 independent frames from each calibration phantom. The second procedure, termed stable acquisition, involved securing the transducer using a bar clamp holder. Subsequently, ten identical frames were captured using both the SonixOne and Verasonics machines from precisely the same position on the calibration phantoms. These procedures facilitated the acquisition of calibration data necessary for computing the M2M transfer function.

As imaging settings, line by line acquisition with 2 cm axial focal point was used for both machines. In the SonixOne, the center pulse frequency was set at 9 MHz and its output power level was set at 0 dB. In the Verasonics, the center pulse frequency was set at 5 MHz and its output power level was set at 45.2 Voltage. These settings were configured to evaluate the proposed method under combined hardware and acquisition-related mismatches.

\subsection{Data Preparation}

After all processing, an ultrasound image frame size from both the testing and training machines was 2,080 pixels $\times$ 256 pixels after all processing. The axial depth was 4 cm. The DL network utilized the raw backscattered RF data as its input. After obtaining frames, square data patches from the frames were extracted to be employed in the DL network. These patches measured 200 samples $\times$ 26 samples, corresponding to physical dimensions of 4 mm $\times$ 4 mm. The motivation for patch extraction is rooted in traditional QUS approaches. In traditional tissue characterization, a data patch is extracted from the ultrasound image to examine the ultrasound signals within a region of interest.

\begin{figure}[hbt!]
\begingroup
    \centering
    \begin{tabular}{c}
\hspace{-0.0cm}{ \includegraphics[width=0.6\linewidth]{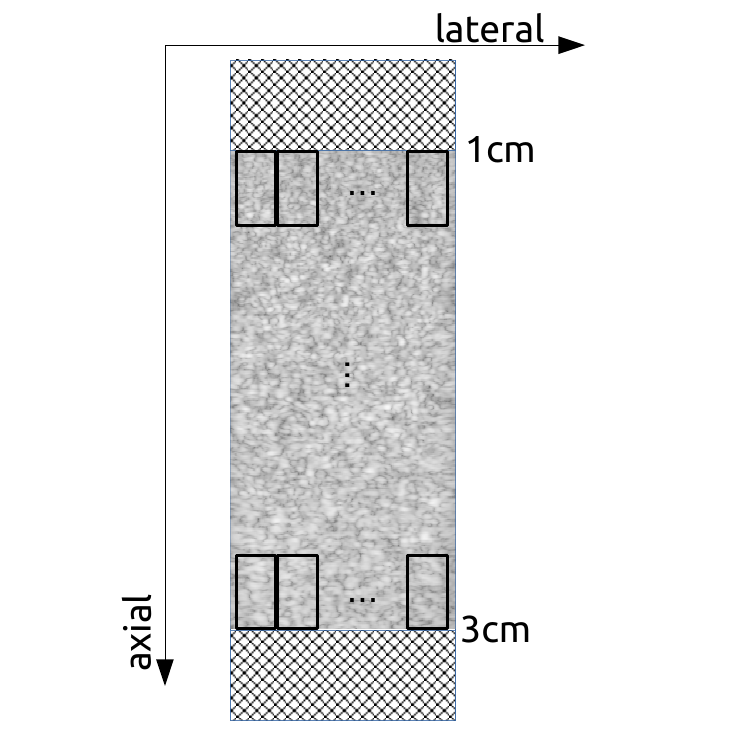}}\\
\end{tabular}
\caption{The Process of Patch Extraction in Data Preparation}
\label{fig:patch_extract}
\endgroup
\end{figure}

We were able to extract 81 image patches (9 lateral positions and 9 axial positions) from a single frame as illustrated in Fig. \ref{fig:patch_extract}. During the extraction process, the initial 540 pixels were omitted. The next sequence of patches in the lateral direction was generated by moving the beginning of the succeeding patch by 26 pixels. In the axial direction, the next sequence of patches was generated by moving the beginning of the succeeding patch by 100 pixels. This method resulted in 9 patches axially by 9 patches laterally, allowing us to extract a total of 81 image patches from each ultrasound image.

From the training machine, 2000 ultrasound frames were acquired, with 1000 frames from each classification phantom, to be used in training, resulting in 162,000 patches. From the testing machine, 1000 ultrasound frames were acquired, with 500 frames from each classification phantom, to be used in testing, leading to 81,000 patches. Regarding calibration data, through stable acquisition, 10 frames from a fixed point were acquired for each machine, and through free-hand acquisition, 1000 frames were acquired for each machine.

\subsection{Calibration}

In a prior work, a transfer function was developed to mitigate acquisition-related data mismatches within the same imaging machine \cite{soylu2023calibrating}. In this instance, data mismatch occurs at the machine level, including acquisition and hardware-level mismatches. We follow an identical derivation and notation to develop the M2M transfer function. A simplified decomposition of an ultrasound image involves the system response and the tissue signal.

\begin{align}
\label{powerspec_simple}
    I(x, f) = S_\phi(x, f) P(x, f)
\end{align}
The system response, denoted by $S_\phi$, holds the information associated with the ultrasound imaging system and the tissue signal, denoted by $P$,  holds the information associated with the imaging substrate. To mitigate the mismatches between two machines $\phi_{train}$ and $\phi_{test}$, we used a calibration phantom $P$, such that,
\begin{align}
\label{powerspec}
   \frac{I_{test}(x, f)}{I_{train}(x, f)} &= \frac{S_{\phi_{test}}(x, f)}{S_{\phi_{train}}(x, f)}\\ &= \Gamma_{M2M}(x, f)
\end{align}
The M2M transfer function, denoted by $\Gamma_{M2M}$, is capable of transferring between training and testing machines.

To calibrate a DL network in training time, i.e. train-time calibration (Algorithm \ref{alg:algo_train}),  $\Gamma_{M2M}$ can be used:
\begin{align}
\label{training time}
    I_{test}(x, f) &= \Gamma_{M2M}(x, f) I_{train}(x, f),   
\end{align}
$\Gamma_{M2M}$ transformed the training domain into the testing domain. Following that, the model was trained at the test domain directly. This process was referred to as train-time calibration. On the other hand, $\Gamma_{M2M}^{-1}$ can be used for calibrating DL in testing time, i.e. test-time calibration (Algorithm \ref{alg:algo_test}).
\begin{align}
\label{test time}
    I_{train}(x, f) &= \Gamma_{M2M}^{-1}(x, f) I_{test}(x, f), 
\end{align}
Test-time calibration means that the test set is attempted to be transformed into the training domain through the M2M transfer function so that the originally trained model can be used on the test dataset. Note that the test-time calibration is quicker to implement because a new model does not need to be trained.

In this work, we investigated two methods for calculating the M2M transfer function. One caveat is that we used the same array probe, i.e., the L9-4/38 transducer, but attached to the different systems. In the first method, stable acquisition was implemented. This involved fixing the transducer using holders and clamps. Following the acquisition of calibration data from one machine, the probe seamlessly transitioned to the other machine without altering its position on the calibration phantom by simply moving the connector from one machine to the other. In the second method, free-hand acquisition was used, involving free-hand motion to record a video of 1000 ultrasound frames from the calibration phantom using both testing and training machines. Additionally, two different types of calibration phantoms, which have uniform scattering properties, were utilized to investigate the importance of calibration phantom selection.

Implementation details of the M2M transfer function are identical to the previous work \cite{soylu2023calibrating}. The approach taken to incorporate the M2M transfer function drew inspiration from the Wiener filter \cite{lim1979enhancement},
\begin{align}
\label{wiener train}
    \Gamma^{Wiener} &= \frac{|\Gamma|^{-1}}{|\Gamma|^{-2}+SNR^{-1}}.
\end{align}
This offers a robust method, utilizing the entire spectrum, producing a smoother M2M transfer function. For simplicity, $\Gamma_{M2M}$ and $\Gamma_{M2M}^{-1}$ represent the Wiener filter implementation for the rest of the paper. Furthermore, the M2M transfer function was computed at different depths to accommodate variations in behavior across the depth range. The calibration techniques are explained in greater detail in Algorithm \ref{alg:algo_train} and Algorithm \ref{alg:algo_test}.

\begin{algorithm}[hbt!]
\caption{Train-Time Calibration}\label{alg:algo_train}
\KwData{Training Data: \{$X_{train}$, $y_{train}$\}, Testing Data: \{$X_{test}$, $y_{test}$\}, Calibration Data: \{$X_{calibration}$\}}
\kwPreparation{Calculate $\Gamma_{M2M}$ \& Normalize $X_{train}$}
 Step1: $W_{train}$\&$W_{test}$  $\xleftarrow{}$ $\mathcal{F}\{X_{calibration}\}$\;
 Step2: $\Gamma_{M2M}$ $\xleftarrow{}$ $W_{train}$\&$W_{test}$\;
 Step3: $X_{train}$, $X_{valid}$ = split($X_{train}$)\;
 Step4: $X_{train} \xleftarrow{} \Gamma_{M2M}\{X_{train}\}$\;
 Step5: $X_{valid} \xleftarrow{} \Gamma_{M2M}\{X_{valid}\}$\;
 Step6: $X_{train} \xleftarrow{} (X_{train} - \mathbb{E}X_{train})/\sigma X_{train}$\;
 Step7: $X_{valid} \xleftarrow{} (X_{valid} - \mathbb{E}X_{valid})/\sigma X_{valid}$\;
\kwTraining{Launch a DL model $f_{\theta}$}
\While{epoch}{
Update parameters in $f_{\theta}$ \;
}
\kwInference{Measure Performance} 
Step1: Z-score Normalization $X_{test}$\;
Step2: $y = f_{\theta}(X_{test})$ \;
Step3: Calculate the metrics using $y$ and $y_{test}$ \;
\end{algorithm}

\begin{algorithm}[hbt!]
\caption{Test-Time Calibration}\label{alg:algo_test}
\KwData{Training Data: \{$X_{train}$, $y_{train}$\}, Testing Data: \{$X_{test}$, $y_{test}$\}, Calibration Data: \{$X_{calibration}$\}}
\kwPreparation{Calculate $\Gamma^{-1}_{M2M}$ \& Normalize $X_{train}$}
 Step1: $W_{train}$\&$W_{test}$  $\xleftarrow{}$ $\mathcal{F}\{X_{calibration}\}$\;
 Step2: $\Gamma^{-1}_{M2M}$  $\xleftarrow{}$ $W_{train}$\&$W_{test}$\;
 Step3: $X_{train}$, $X_{valid}$ = split($X_{train}$)\;
 Step6: $X_{train} \xleftarrow{} (X_{train} - \mathbb{E}X_{train})/\sigma X_{train}$\;
 Step7: $X_{valid} \xleftarrow{} (X_{valid} - \mathbb{E}X_{valid})/\sigma X_{valid}$\;
\kwTraining{Launch a DL model $f_{\theta}$}
\While{epoch}{
Update parameters in $f_{\theta}$\;
}
\kwInference{Measure Performance} 
 Step1: $X_{test} \xleftarrow{} \Gamma^{-1}_{M2M}\{X_{test}\}$ \;
 Step2: Z-score Normalization $X_{test}$\;
 Step3: $y = f_{\theta}(X_{test})$ \;
 Step4: Calculate the metrics using $y$ and $y_{test}$ \;
\end{algorithm}

\subsection{Training}
The DL algorithms were trained utilizing a workstation equipped with four NVIDIA RTX A4000. Each experiment was conducted using all four RTX A4000s in parallel. The PyTorch library \cite{paszke2019pytorch} was utilized for all experiments. 

In all experiments, we utilized the Adam algorithm \cite{kingma2014adam} as the optimizer. Hyper-parameters, including epoch numbers and learning rates, were determined aiming for “asymptotic test accuracy". The batch size was selected as 6144 for ResNet experiments and 2048 for DenseNet experiments to maximize memory utilization. During training, a standard method for data augmentation involved applying a horizontal flip with a 50\% probability by default. As training loss, cross-entropy loss was utilized. Z-score normalization/standardization was carried out at the patch level as a data preprocessing step. This process includes subtracting the mean patch, then dividing by the standard deviation patch.

Each experiment, i.e., the training, was repeated 10 times. Next, the average of the classification accuracies, the average area under the receiver operator characteristic(ROC) curve (AUC) and their respective standard deviations were computed using the test sets. The results were obtained patch-wise. The variance in the results was caused by the random initialization of network parameters at each repetition. In the code, random seed was included, ensuring that the results were reproducible.

\subsection{Network Structure}

We employed two established CNN architectures in this study: ResNet-50 \cite{he2016deep} and DenseNet-201 \cite{huang2017densely}. We made minor adjustments to the CNN architectures to customize their input-output relationship to suit our specific problem.. The first convolutional layers, which originally took three input feature channels, were replaced with a single input-channel convolution layer. Additionally, the last layer, a fully connected layer, was also modified to output a single probability corresponding to two classes. For network parameter initialization, pretrained weights were used, except for the first convolutional layer and the last fully connected layer, which were initialized using the default method in PyTorch. During training, all the parameters were unfrozen and fine-tuned through backpropagation.

\section{Results}
\label{sec:results}

\subsection{Train vs Test Statistics}
We compare train-time calibration (Algorithm \ref{alg:algo_train}), test-time calibration (Algorithm \ref{alg:algo_test}), and no calibration cases in Table \ref{tab:res1}. The 'no calibration' experiment represents the scenario where no calibration was implemented and can be thought of as Algorithm \ref{alg:algo_train} without steps 4 and 5 in the preparation section. We set the learning rate to $1e-5$ and the number of epochs to 50 for both train-time calibration and test-time calibration. For no calibration, we set the learning rate to $5e-6$, and we ran the training for 25 epochs. For each algorithm, three different z-score normalizations were implemented at inference time, which corresponds to step 1 in Algorithm \ref{alg:algo_train} and step 2 in Algorithm \ref{alg:algo_test}. For each experiment, z-score normalization strategies at inference can be found in Table \ref{tab:res1statistics} and \ref{tab:res1statisticstest}. In these experiments, stable acquisition with Calibration Phantom 1 was utilized to obtain the M2M transfer function. The results reveal the importance of statistics in z-score normalization. Any shift in those statistics could lead to significantly lower performance.

\begin{table}[htb]
    \centering
\caption{Different Z-Score Normalization at Inference for Train-Time Calibration and No Calibration}
{\small
\begin{tabular}{|p{1.4cm}||p{5cm}|}
 \hline
\textit{Statistics Type}& \textit{Normalization} \\
 \hline
Train Statistics & $X_{test} = \frac{X_{test} - \mathbb{E}X_{train}}{\sigma X_{train}}$ \\
 \hline
Test Statistics&  $X_{test} = \frac{X_{test} - \mathbb{E}X_{test}}{\sigma X_{test}}$  \\
 \hline
Calibrated Statistics& $X_{test} = \frac{X_{test} - \mathbb{E}\Gamma_{M2M}\{X_{train}\}}{\sigma \Gamma_{M2M}\{X_{train}\}}$ \\
 \hline
\end{tabular}
}
\label{tab:res1statistics}
\end{table}

\begin{table}[htb]
    \centering
\caption{Different Z-Score Normalization at Inference for Test-Time Calibration}
{\small
\begin{tabular}{|p{1.4cm}||p{5.5cm}|}
 \hline
\textit{Statistics Type}& \textit{Normalization} \\
 \hline
Train Statistics & $X_{test} = \frac{ \Gamma_{M2M}^{-1} \{X_{test}\} - \mathbb{E}X_{train}}{\sigma X_{train}}$ \\
 \hline
Test Statistics& $X_{test} = \frac{ \Gamma_{M2M}^{-1} \{X_{test}\} - \mathbb{E}X_{test}}{\sigma X_{test}}$ \\
 \hline
Calibrated Statistics& $X_{test} = \frac{\Gamma_{M2M}^{-1} \{X_{test}\} - \mathbb{E}\Gamma_{M2M}^{-1}\{X_{test}\}}{\sigma \Gamma_{M2M}^{-1}\{X_{test}\}}$ \\
 \hline
\end{tabular}
}
\label{tab:res1statisticstest}
\end{table}
\begin{table*}[t]
    \centering
\caption{Train vs Test Statistics}
{\renewcommand{\arraystretch}{1.2}
{\small
\begin{tabular}{ |p{0.25cm}||p{3.7cm}||p{2.5cm}||p{2.8cm}||p{2.5cm}||p{2.5cm}|}
 \hline
\textit{No}&\textit{Experiment}&\textit{ResNet-Acc.}&\textit{DenseNet-Acc.}&\textit{ResNet-AUC}&\textit{DenseNet-AUC}\\
 \hline
1&Train-Time Calibration with Train Statistics & 50.02$\pm$0.01 & 49.54$\pm$1.07 & 0.360$\pm$0.123 & 0.462$\pm$0.091\\
 \hline
2&Train-Time Calibration with Calibrated Statistics & 86.06 $\pm$3.95 & 81.17$\pm$5.52 & 0.985$\pm$0.005 & 0.985$\pm$0.008\\
 \hline
3&Train-Time Calibration with Test Statistics& 88.42$\pm$4.20 & 84.41$\pm$4.80 & 0.988$\pm$0.007 & 0.986$\pm$0.008\\
 \hline
4&Test-Time Calibration with Train Statistics& 80.80$\pm$7.83 & 74.86$\pm$8.08 & 0.995$\pm$0.006 & 0.987$\pm$0.011\\
 \hline
5&Test-Time Calibration with Calibrated Statistics& 80.76$\pm$6.87 & 76.01$\pm$5.50 & 0.994$\pm$0.007 & 0.988$\pm$0.009\\
 \hline
6&Test-Time Calibration with Test Statistics& 50.12$\pm$4.04 & 54.46$\pm$7.04 & 0.577$\pm$0.188 & 0.629$\pm$0.109\\
 \hline
7&No Calibration with Train Statistics& 50.01$\pm$0.01 & 47.19$\pm$5.94 & 0.405$\pm$0.073 & 0.488$\pm$0.157\\
\hline
8&No Calibration with Calibrated Statistics& 70.59$\pm$4.73 &  68.40$\pm$2.86 & 0.936$\pm$0.021 & 0.919$\pm$0.017\\
\hline
9&No Calibration with Test Statistics& 75.11$\pm$4.67 & 72.65$\pm$3.23 & 0.949$\pm$0.011 & 0.936$\pm$0.139\\
\hline
\end{tabular}
}}
\label{tab:res1}
\end{table*}

\subsection{Different Calibration Phantoms}

We investigated the effects of using different calibration phantoms, Calibration Phantom 1 and Calibration Phantom 2, on the success of calibration, as shown in Table \ref{tab:res2}. For train-time calibration and test-time calibration, we set the learning rate to $1e-5$, and ran the training for 50 epochs. Stable acquisition was utilized in these results. The results reveal the importance of calibration phantom selection. For train-time calibration and test-time, utilizing different calibration phantoms led to different behaviour. 

\begin{table*}[t]
    \centering
\caption{Different Calibration Phantoms}
{\renewcommand{\arraystretch}{1.2}
{\small
\begin{tabular}{ |p{0.25cm}||p{3.7cm}||p{2.5cm}||p{2.8cm}||p{2.5cm}||p{2.5cm}|}
 \hline
\textit{No}&\textit{Experiment}&\textit{ResNet-Acc.}&\textit{DenseNet-Acc.}&\textit{ResNet-AUC}&\textit{DenseNet-AUC}\\
 \hline
1&Train-Time Calibration with Calibration Phantom 1& 88.42$\pm$4.20 & 84.41$\pm$4.80 & 0.988$\pm$0.007 & 0.986$\pm$0.008\\
 \hline
2&Train-Time Calibration with Calibration Phantom 2& 84.02$\pm$5.29 & 78.91$\pm$5.07 & 0.989$\pm$0.005 & 0.986$\pm$0.006\\
 \hline
3&Test-Time Calibration with Calibration Phantom 1& 80.76$\pm$6.87 & 76.01$\pm$5.50 & 0.994$\pm$0.007 & 0.988$\pm$0.009\\
 \hline
4&Test-Time Calibration with Calibration Phantom 2& 81.95$\pm$6.41 & 77.51$\pm$5.36 & 0.993$\pm$0.006 & 0.987 $\pm$0.008\\
 \hline
\end{tabular}
}}
\label{tab:res2}
\end{table*}

\subsection{Stable vs Hand-Free Calibration}

We investigated the effects of using different acquisitions for the calibration data, stable and free-hand, on the success of calibration, as shown in Table \ref{tab:res3}. For train-time calibration and test-time calibration, we set the learning rate to $1e-5$, and ran the training for 50 epochs. Calibration Phantom 1 was utilized in these results. The results reveal that stable and free-hand acquisition led to similar performances, with free-hand providing slightly better accuracies.

\begin{table*}[t]
    \centering
\caption{Stable vs Hand-Free Calibration}
{\renewcommand{\arraystretch}{1.2}
{\small
\begin{tabular}{ |p{0.25cm}||p{3.7cm}||p{2.5cm}||p{2.8cm}||p{2.5cm}||p{2.5cm}|}
 \hline
\textit{No}&\textit{Experiment}&\textit{ResNet-Acc.}&\textit{DenseNet-Acc.}&\textit{ResNet-AUC}&\textit{DenseNet-AUC}\\
 \hline
1&Train-Time Calibration with Stable Calibration& 88.42$\pm$4.20 & 84.41$\pm$4.80 & 0.988$\pm$0.007 & 0.986$\pm$0.008\\
 \hline
2&Train-Time Calibration with Hand-Free Calibration& 88.51$\pm$3.56 & 85.13$\pm$4.70 & 0.986$\pm$0.007 & 0.986$\pm$0.007\\
 \hline
3&Test-Time Calibration with Stable Calibration& 80.76$\pm$6.87 & 76.01$\pm$5.50 & 0.994$\pm$0.007 & 0.988$\pm$0.009\\
 \hline
4&Test-Time Calibration with Hand-Free Calibration& 81.28$\pm$6.91 & 76.63$\pm$5.59 & 0.994$\pm$0.006 & 0.989$\pm$ 0.009\\
 \hline
\end{tabular}
}}
\label{tab:res3}
\end{table*}

\section{Discussion}
\label{sec:discussion}

The M2M transfer function has the potential to be implemented in practice as it does not rely on the acquisition of test domain samples to calibrate the classifier. The approach can provide a practical means to transfer DL models between imaging machines and to transfer data from different sources to the desired domain, thereby significantly reducing model development costs. In this article, a M2M transfer function was investigated by utilizing different normalization strategies at inference, different calibration phantoms, and different acquisition strategies for acquiring calibration data. We observed that the M2M transfer function was effective in calibrating a DL model between imaging machines, increasing mean classification accuracy from 50.01\% to 88.51\% and mean AUC from 0.405 to 0.995. 

In Table \ref{tab:res1}, we mainly observe that the M2M transfer function significantly improved accuracy and AUC under machine-level data mismatches. Specifically, the use of test statistics as oracle information boosted performance to the highest level. Additionally, there were multiple interesting observations that can be derived from the table. First, in the case of no calibration, while the use of training statistics resulted in very poor performance, the utilization of calibrated statistics and oracle statistics led to a significant improvement in accuracy and AUC. The accuracy improved from 50\% to the range of 70-75\%, and the AUC increased from 0.5 to above 0.9. It is worth noting that a significant improvement was achieved by using calibrated statistics even without calibrating input data. This improvement was observed solely by implementing calibration for the statistics used in the normalization step, demonstrating the potential of the M2M transfer function. The results from experiment 3 indicated that in addition to calibrating normalization statistics, when input data were also calibrated, the accuracy reached around 90\%, while the AUC reached levels around 0.99. On the other hand, the results from experiment 1 indicated that shifts in z-score normalization statistics could lead to a drastic drop in performance, even when the input data was calibrated. This highlights the importance of the statistics used during calibration in the case of data mismatch. Another important observation from Table \ref{tab:res1} is that the train-time calibration was significantly more successful than test-time calibration in terms of accuracy. However, in terms of AUC, they were comparable, with test-time calibration actually being slightly better. This result indicates the potential for test-time calibration to achieve a similar level of accuracy performance as train-time calibration. However, further study is needed to develop a systematic approach to enhance the test-time calibration process, which should include better strategies for hyperparameter tuning. In terms of network architecture, ResNet slightly outperformed DenseNet after calibration. One potential explanation for this difference is that ResNet was 50 layers deep, whereas DenseNet was 201 layers deep, which aligns with the common understanding that increasing depth and batch normalization layers can make the calibration process more challenging.

In Table \ref{tab:res2}, Calibration Phantom 1 and Calibration Phantom 2 were used for both train-time and test-time calibration. We observed that Calibration Phantom 1 resulted in better calibration for train-time calibration, while for test-time calibration, Calibration Phantom 2 performed slightly better in terms of accuracy. In terms of AUC, both calibration phantoms yielded similar results. Overall, these results suggest that the selection of a calibration phantom was relevant to performance and there is an intriguing relationship between the selection of the calibration phantom and performance. Even though the proposed method did not rely on the acquisition of samples from the test domain, one could hypothesize that the calibration phantom selection should align with the classification samples. Intuitively, the properties of the calibration phantom should resemble those of the test and training domain samples to enhance the calibration process. However, further studies should be conducted to develop a systematic approach for selecting a calibration phantom based on properties of the training domain, which is known.

In Table \ref{tab:res3}, stable acquisition and free-hand acquisition were investigated in terms of calibration performance. In stable acquisition, the M2M transfer function was calculated using a single fixed view from the calibration phantom. In free-hand acquisition, a video of ultrasound frames was recorded, and the M2M transfer function was calculated by averaging it over the frames. The results indicate that stable and free-hand acquisition led to similar performances, with free-hand providing slightly better accuracies. This may sound counter intuitive at first as free-hand acquisition uses more frames to calculate a M2M transfer function; however, this observation actually verifies the robustness of the Wiener-inspired implementation against noise. Apparently, the Wiener-inspired implementation provides a robust method to calibrate data mismatches using just a single, fixed ultrasound frame. That being said, as a future direction, utilizing multiple calibration views to enhance calibration performance still remains an attractive avenue.

Overall, the results indicate that using oracle information for z-score normalization at inference results in performance improvement compared to using non-oracle statistics. As a side note, oracle statistics refer to the true statistics from the test domain, while non-oracle statistics refer to any statistics that can be derived from training data and/or utilizing M2M transfer function. Even though during test-time calibration, this observation was not as pronounced because test-time statistics could not be utilized directly. In Table \ref{tab:res1}, Experiment 4, which used non-oracle statistics (train statistics in this case), and Experiment 5, which used oracle statistics (calibrated statistics using test data), performed very similarly. For train-time calibration, this observation is clear. Experiment 3, which used oracle statistics (test statistics in this case), performed better than Experiment 2, which used non-oracle statistics (calibrated statistics using training data). Although we presented the primary advantage of the M2M transfer function as not relying on the acquisition of samples from the test domain, this may sound contradictory. However, using oracle statistics was still a less strict requirement than acquiring the test domain data itself. Another point regarding the results is that there was generally a high level of variance. This result was a consequence of the assumption of not having access to test domain sample data and imperfect calibration. On the other hand, the high variance indicates that if the test domain sample data were available, it could be used in validation to almost perfectly calibrate the DL model, similar to the case where test domain data was accessible during training. 

The results of this work highlight several potential future directions. First, the results indicate that the selection of a calibration phantom can affect performance. Therefore, developing a systematic procedure for choosing a calibration phantom remains an important problem. Second, enhancing calibration performance by leveraging multiple calibration views could provide additional benefit. Third, the impact of using different transducers on calibration and devising solutions to address potential challenges arising from variations in transducer bandwidth requires additional study. Similarly, the effects of different acquisition techniques, such as plane wave imaging versus line by line imaging or even changes in sampling rate, on the calibration may also affect the ability to transfer classification models from one machine to another. From a security perspective, in cases where DL model transferability is not desired, it may be possible to develop defense mechanisms, such as using sampling rate mismatches. Moreover, if data acquired from multiple machines can be combined through a M2M transfer function, the increase in data availability, i.e., incorporating the data from multiple machines, could lead to improved DL models. The code for the implementation of training, testing, and calibration can be accessed at the following repository: https://github.com/usoylu2/m2m. The dataset is available for use via the following link: https://uofi.box.com/s/d9ecw002ree6gj9tlplz7t0i2f1ojbk7.

\section{Conclusion}
\label{sec:conclusion}
We introduced a M2M transfer function for mitigating the effects of data mismatches between data acquired from different ultrasound scanners. The results indicate that the M2M transfer function can be effective in calibrating mismatches between different imaging machines. Therefore, the incorporation of M2M transfer function can offer an economical approach to transferring datasets and DL models between machines, reducing the cost of model development and paving the way for an enhanced understanding of model security.
\bibliographystyle{IEEE_ECE}
\bibliography{refs}
\end{document}